\documentclass
	[a4paper,singlecolumn,twoside,12pt,showkeys]
	{revtex4}

\usepackage{amsmath,amssymb}	
\usepackage{bm}					
\usepackage{color}					
\usepackage{graphics}				
\usepackage{dcolumn}				
\usepackage[colorlinks=true, citebordercolor={0 0 0}, linkcolor=blue, citecolor=blue, urlcolor=blue]{hyperref}
\usepackage{setspace}
\usepackage{ulem}

\makeatletter
\def\frontmatter@abstractwidth{0.9\textwidth}	
\makeatother

\begin{document}


\newcommand{\By}{$\times$}
\newcommand{\SqrtBy}[2]{$\sqrt{#1}$\kern0.2ex$\times$\kern-0.2ex$\sqrt{#2}$}
\newcommand{\Degree}{$^\circ$}
\newcommand{\DegreeC}{$^\circ$C}
\newcommand{\Ohmcm}{$\Omega\cdot$cm}

\title{
Wave function of a photoelectron and its collapse in the photoemission process 
}


\author{Hiroaki Tanaka}
\email[Corresponding author: ]{hiroaki-tanaka@issp.u-tokyo.ac.jp}
\affiliation{%
Institute for Solid State Physics, The University of Tokyo, Kashiwa, Chiba 277-8581, Japan
}

\begin{abstract}

\vspace*{1mm}
Based on the first-order perturbation theory, we show that the wave function of a photoelectron is a wave packet with the same width as the incident light pulse.
Photoelectron detection measurements revealed that the widths of signal pulses were much shorter than the light pulse and independent of the origin (photoemission or other noises), which is an experimental observation of the wave function collapse.
Signal pulses of photoelectrons were distributed along the time axis within the same width as the light pulse, consistent with the interpretation of a wave function as a probability distribution.
\begin{center}
\includegraphics{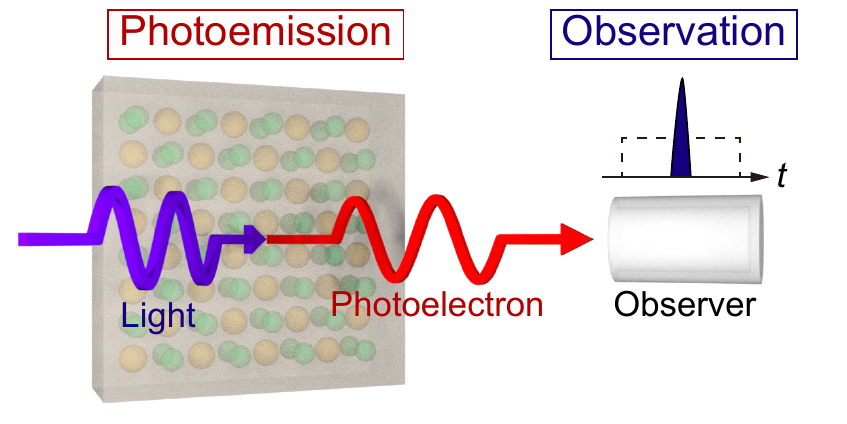}
\end{center}
\end{abstract}


\maketitle
\newpage

The photoelectric effect, the emission of electrons due to the irradiation of ultraviolet light or x-rays, was found in 1887 by Hertz \cite{Hertz1887,Hufner2003}.
The kinetic energy of the liberated electrons and the light frequency dependence of the phenomenon cannot be explained by classical physics, and the photoelectric effect, as well as the black body radiation and the Compton scattering, invoked the development of quantum physics \cite{Messiah1964}.
The photoemission process has been utilized to investigate the electronic structure of solids in the name of photoemission spectroscopy (PES) and angle-resolved PES (ARPES).
Since momenta and energies of photoelectrons are related to Bloch wavevectors and binding energies in crystals, ARPES can experimentally determine the band dispersion of solids \cite{RevModPhys.93.025006}.
Owing to the surface sensitivity of ARPES \cite{Seah1970}, this technique has been utilized to investigate the electronic structures of two-dimensional materials and crystal surfaces such as graphene \cite{PhysRevB.51.13614,PhysRevB.77.155303} and topological surface states \cite{Xia2009}.

The photoemission process has been described by the first-order perturbation theory and the one-electron approximation \cite{Hufner2003, RevModPhys.93.025006}.
The perturbation theory treats the incident light as a perturbative vector potential, and the energy conservation law is derived from Fermi's golden rule.
The matrix element term in the equation of the excitation probability gives momentum conservation law.
Here, the final state, the wave function of a photoelectron, is a plane wave in the vacuum and rapidly decays into the bulk in the one-step model \cite{MOSER201729,RevModPhys.93.025006}; the photoelectron wave function infinitely spreads in the vacuum.
On the other hand, the three-step model, more widely used than the one-step model, describes a photoelectron as a classical particle generated in the bulk and escaping from the surface \cite{RevModPhys.93.025006}.
Furthermore, time-of-flight spectrometers measure the flight time of a photoelectron to determine its kinetic energy \cite{OVSYANNIKOV201392}, where the photoelectron is treated as a particle.
A photoelectron has been depicted as both a wave and a particle depending on models and the concrete description has been missing.

In this report, we discuss the time evolution of a photoelectron wave function based on the time-dependent first-order perturbation theory.
We derive that the wave function is the same as a plane wave only in the area determined by the incident light position, the momentum of the photoelectron, and the time, which is consistent with classical intuition.
The time width of the wave packet is the same as the pulse width of the incident light.
We experimentally generated 80 $\mu\mathrm{s}$-long wave packets using a single crystal of FeSe and a 4.66 eV CW laser and detected them by a channel electron multiplier (CEM).
The photoelectron signal was shorter than 10 ns, which means that the wave function indeed collapses in the observation process by the CEM.
We also demonstrate that wave functions are interpreted as probability distributions from the spread of photoelectron signal positions.

\begin{figure}
\includegraphics{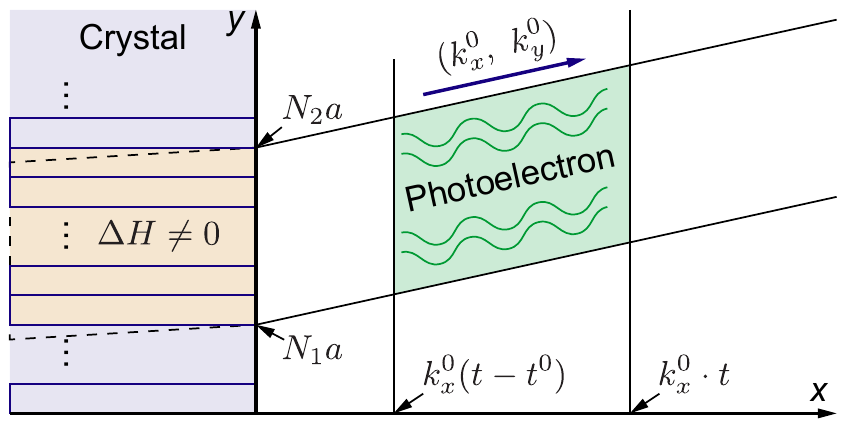}
\caption{\label{Fig: Wavefunction} Propagation of a photoelectron wave function emitted from a crystal. The terms in the figure are defined in the main text.}
\end{figure}

We consider a system like Figure \ref{Fig: Wavefunction} to calculate a photoelectron wave function; A crystal is placed in the $x<0$ area, the incident light with frequency $\omega$ is irradiated in the $N_1a<y<N_2a$ region and from $t=0$ to $t=t^0$, where $a$ is the lattice constant of the crystal along the $y$ direction, $N_1$ and $N_2$ are integers.
The nonzero perturbation area is drawn in orange in Figure \ref{Fig: Wavefunction}.
If the incident light is not perpendicular to the crystal surface, the area should be tilted like the black dashed parallelogram.
However, it can be approximated to the orange rectangle because the mean free paths of photoelectrons are few nanometers \cite{Seah1970}, much shorter than the beam spot size $N_2a-N_1a$.
Although the system is two-dimensional, the extension of the argument to the three-dimensional case is straightforward.
We take an initial state $|\psi_{(m_x, k_y^0)}^\mathrm{I}\rangle$ as an eigenstate of the Hamiltonian and the translation along the $y$ direction; since the crystal structure is terminated at $x=0$, we do not use the translational symmetry along the $x$ direction.
$k_y^0$ is the Bloch wavevector along the $y$ direction, and $m_x$ is an index to distinguish other degenerations.
$\omega^0$ represents the eigenenergy of the initial state; hereafter, we use the atomic unit, where $\hbar$ (Dirac constant), $m$ (electron mass), $a_0$ (Bohr radius), and $E_h$ (Hartree energy) are set to unity.
For the final states we use wave functions which are plane waves in the vacuum $|\psi_{(k_x,k_y)}^\mathrm{F}\rangle=e^{\mathrm{i}\mathbf{k}\cdot\mathbf{r}}$, where $\mathbf{k}$ and $\mathbf{r}$ represent $(k_x,\ k_y)$ and $(x,\ y)$, respectively, and the eigenenergy $\omega_{(k_x,k_y)}$ is equal to $\frac{1}{2}(k_x^2+k_y^2)$.
Since the wave function of the final state in the crystal is unimportant for the discussion, we just assume it to be localized on the surface consistent with the very short mean free paths of photoelectrons and to satisfy the boundary condition with the plane wave in the vacuum.

From the time-dependent first-order perturbation theory, the excited part of the wave function can be represented by
\begin{gather}
|\psi_{\mathrm{ex}}(t)\rangle=\int \mathrm{d}k_x\mathrm{d}k_y\ c_{(k_x,k_y)}(t^0)|\psi_{(k_x,k_y)}^\mathrm{F}\rangle e^{-\mathrm{i}\omega_{(k_x,k_y)}t}\\
c_{(k_x,k_y)}(t^0)=\frac{1-e^{\mathrm{i}(\omega_{(k_x,k_y)}-\omega^0-\omega)t^0}}{\omega_{(k_x,k_y)}-\omega^0-\omega}\langle \psi_{(k_x,k_y)}^\mathrm{F}|\Delta H|\psi_{(m,k_y^0)}^\mathrm{I}\rangle.
\end{gather}
The fraction part in the coefficient takes a negligible value if $\omega_{(k_x,k_y)}-\omega^0-\omega$ is far from zero; the norm of the fraction becomes Dirac's delta function $\delta(\omega_{(k_x,k_y)}-\omega^0-\omega)$ in the $t^0\rightarrow\infty$ limit and gives the energy conservation law.
We define $k_x^0$ by the rule $\omega_{(k_x^0,k_y^0)}-\omega^0-\omega=0$ and introduce $\delta_x$ and $\delta_y$ as $k_x=k_x^0+\delta_x$ and $k_y=k_y^0+\delta_y$, respectively, then we can neglect the higher-order terms of $\delta_x$ and $\delta_y$ like $\omega_{(k_x,k_y)}-\omega^0-\omega\simeq k_x^0\delta_x+k_y^0\delta_y$.
For the matrix element part, we define $M_0$ as the matrix element when $N_1=0$ and $N_2=1$.
Using Bloch's theorem, we get
\begin{align}
\langle \psi_{(k_x,k_y)}^\mathrm{F}|\Delta H|\psi_{(m,k_y^0)}^\mathrm{I}\rangle&=\sum_{N=N_1}^{N_2-1} e^{-\mathrm{i}k_y Na}e^{\mathrm{i}k_y^0 Na}M_0 \label{Eq: matrix_element_Bloch}\\
&=\frac{e^{-\mathrm{i}\delta_y N_2a}-e^{-\mathrm{i}\delta_y N_1a}}{e^{-\mathrm{i}\delta_y a}-1}M_0,
\end{align}
where the exponentials in Equation (\ref{Eq: matrix_element_Bloch}) represent the Bloch phases of the initial and final states.
Assuming $\delta_x$ is much smaller than the Brillouin zone size of the crystal, we neglect the $\delta_x$ dependence of $M_0$.
Combining these results, finally we obtain
\begin{equation}
|\psi_{\mathrm{ex}}(t)\rangle=\int \frac{1-e^{\mathrm{i}(k_x^0\delta_x+k_y^0\delta_y)t^0}}{k_x^0\delta_x+k_y^0\delta_y}\frac{e^{-\mathrm{i}\delta_y N_2 a}-e^{-\mathrm{i}\delta_y N_1 a}}{-\mathrm{i}\delta_y a}M_0
\times e^{\mathrm{i}(\delta_xx+\delta_yy-(k_x^0\delta_x+k_y^0\delta_y)t)}e^{\mathrm{i}(k_x^0 x+k_y^0 y-\omega_{(k_x^0,k_y^0)}t)}\mathrm{d}\delta_x\mathrm{d}\delta_y,
\end{equation}
where the latter part represents the plane wave $e^{\mathrm{i}(\mathbf{k}\cdot\mathbf{r}-\omega_{(k_x,k_y)}t)}$ with the approximations being applied.
The integration is performed using a new variable $\delta_{x^\prime}=\delta_x+\frac{k_y^0}{k_x^0}\delta_y$, the rectangular function $\Theta(x;\ a,\ b)$, which returns 1 if $a<x<b$ and 0 otherwise, and the formula
\begin{equation}
\Theta(x;\ a,\ b)=\frac{\mathrm{i}}{2\pi}\int \frac{1}{k}(e^{-ikb}-e^{-ika})e^{ikx}\mathrm{d}k
\end{equation}
derived from the Fourier transformation.
The result is
\begin{equation}
|\psi_{\mathrm{ex}}(t)\rangle=(\mathrm{const.})\cdot e^{\mathrm{i}(k_x^0 x+k_y^0 y-\omega_{(k_x^0,k_y^0)}t)}\Theta(x;\ k_x^0(t-t_0),\ k_x^0 t) \Theta\left(y;\ N_1a+\frac{k_y^0}{k_x^0} x,\ N_2a+\frac{k_y^0}{k_x^0} x\right).
\end{equation}

The wave packet determined by the rectangular functions is drawn in green in Figure \ref{Fig: Wavefunction}.
This is very natural; the plane wave corresponding to the wavevector $\mathbf{k}=(k_x^0,\ k_y^0)$ satisfies the conservation laws of momentum and energy, the vertical edges move to the right with the velocity $k_x^0$ and correspond to the timing of the perturbation switch, and the top and bottom edges are parallel to the vector $(k_x^0,\ k_y^0)$ and touch the edges of the perturbation area.
The difference between the wave function of a photoelectron and the plane wave comes from the incomplete conservation laws of the energy and momentum due to the finite irradiation time and area.
If the light source is short-pulsed, wave packets have a negligible spread in the space, consistent with the classical particle description of photoelectrons.
Typical ARPES measurements satisfy this condition.
Light sources such as pulsed lasers and synchrotron lights have pulse widths shorter than one nanosecond.
Photoelectron signals amplified by a CEM have a pulse width of a few nanoseconds, as shown later, so the time width of the photoelectron wave function is undetectable.
Furthermore, in most ARPES measurements, photoelectron signals are amplified by a micro-channel plate, a two-dimensional array of electron multipliers.
In this case, instead of multiplied signals, luminescence due to the multiplied signals on a phosphor screen is detected by a CCD camera.
All the information from such a photoelectron analyzer is the time-integrated intensity of the photoelectron signals; the shape of the photoelectron wave function is unnecessary to analyze experimental data.
The experimental limitation may not have encouraged the development of the discussion beyond the classical description.
However, if the pulse width is much longer than that of photoelectron signals, quantum physics is necessary to explain the result presented below.

\begin{figure}
\includegraphics{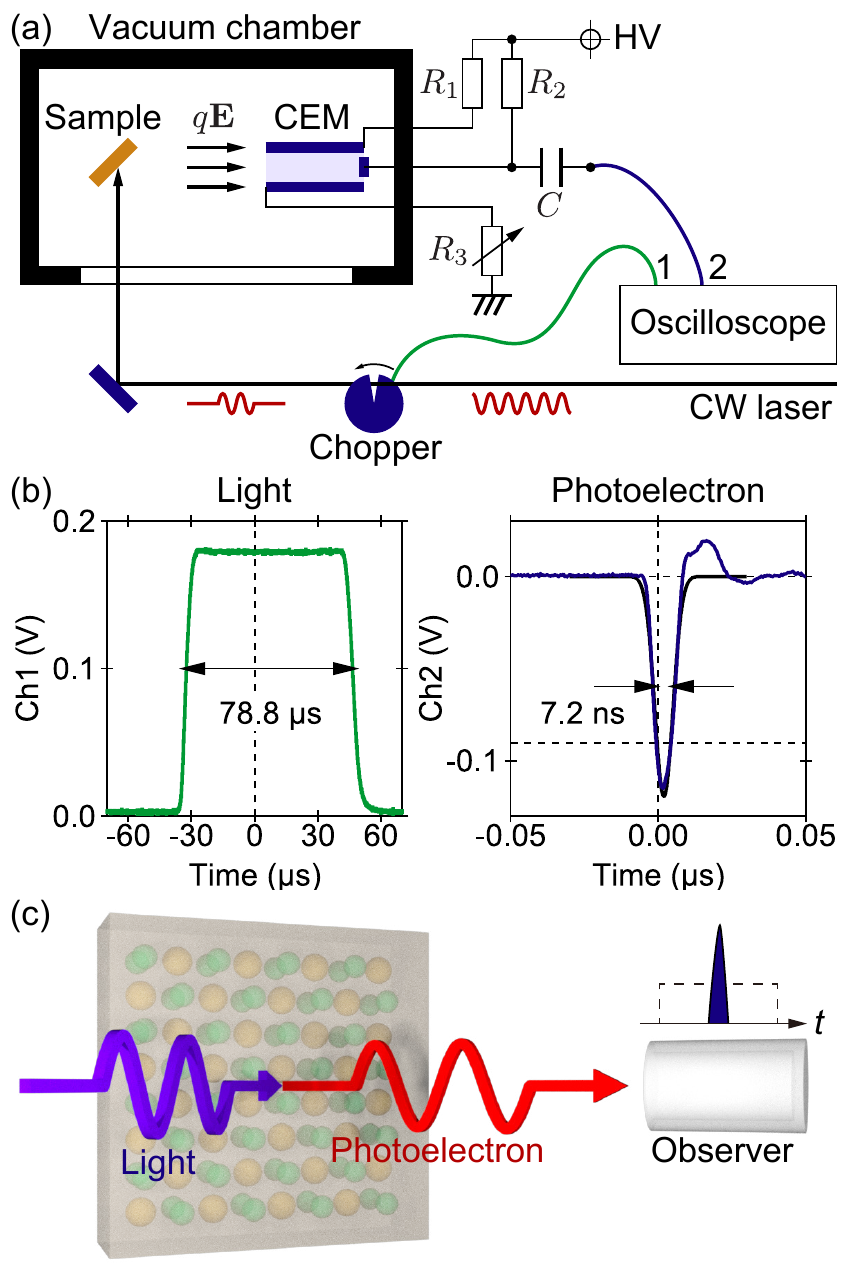}
\caption{\label{Fig: Measurements} Photoelectron detection measurements. (a) Experimental setup. Measurement conditions were $\mathrm{HV}=2.5\ \mathrm{kV},\ R_1=10\ \mathrm{M}\Omega,\ R_2=1\ \mathrm{M}\Omega,\ R_3=5\ \mathrm{M}\Omega,\ C=180\ \mathrm{pF}$. (b) A photoelectron signal (blue curve in the right panel) and the corresponding light pulse signal (green curve in the left panel). In the right panel, the horizontal dashed line is the trigger level and the black curve represents the fitted gauss function to the signal. (c) Schematic of the wave function collapse in the photoemission process.}
\end{figure}

To investigate the behavior of this wave packet, we experimentally generated it and observed it by a CEM.
Figure \ref{Fig: Measurements}(a) shows the setup of the measurements.
We mechanically chopped a 4.66 eV CW laser to generate light pulses longer than those from the CEM (a few nanoseconds).
A single crystal of FeSe and the CEM (Model KBL 5RS from Dr. Sjuts Optotechnik GmbH) was installed in the vacuum chamber.
The combination of the material and the light source is suitable for emitting a sufficient number of photoelectrons; the 4.66 eV CW laser has also been used as a light source for photoemission electron microscopy (PEEM) \cite{Taniuchi2015}, and PEEM measurements of the iron-based superconductor FeSe have successfully discovered nematic states \cite{Shimojima2021}.
The CEM input was applied with a positive voltage to accelerate photoelectrons, as represented by the arrows $q\mathbf{E}$ in Figure \ref{Fig: Measurements}(a), and to increase detection efficiency.
As a result, the difference in the kinetic energies and momenta between photoelectrons just emitted from the surface did not affect the flight time to the CEM and all electrons emitted from the surface reach the CEM irrespective of their momenta.
The FeSe sample was cleaved \textit{in situ} just before the measurements and the light was irradiated on the clean surface.
The laser power was 23 mW ($3.1\times10^{16}\ \textrm{photons/s}$) before the chopper and the duty cycle of the chopper was about 1/63.
The oscilloscope simultaneously measured voltages from the mechanical chopper (Ch1) and the CEM (Ch2); the negative edge trigger at about $-90\ \mathrm{mV}$ was set on Ch2.

Figure \ref{Fig: Measurements}(b) represents one detection event.
The left panel shows that the incident light was turned on for about 80 $\mu\mathrm{s}$, and the corresponding photoelectron wave packet should have the same time width.
However, the photoelectron signal was much narrower than the light pulse [Figure \ref{Fig: Measurements}(b) right panel]; the signal was like a spike well fitted by the gauss function with an FWHM of 7.2 ns.
The difference between the light and photoelectron pulses is due to the principle of the wave function collapse [Figure \ref{Fig: Measurements}(c)] \cite{RevModPhys.76.1267}; when an observer detects a photoelectron via a macroscopic signal, the photoelectron looks like a classical particle located on a certain position.
The wave function gives the probability distribution of where the particle appears.

\begin{figure}
\includegraphics{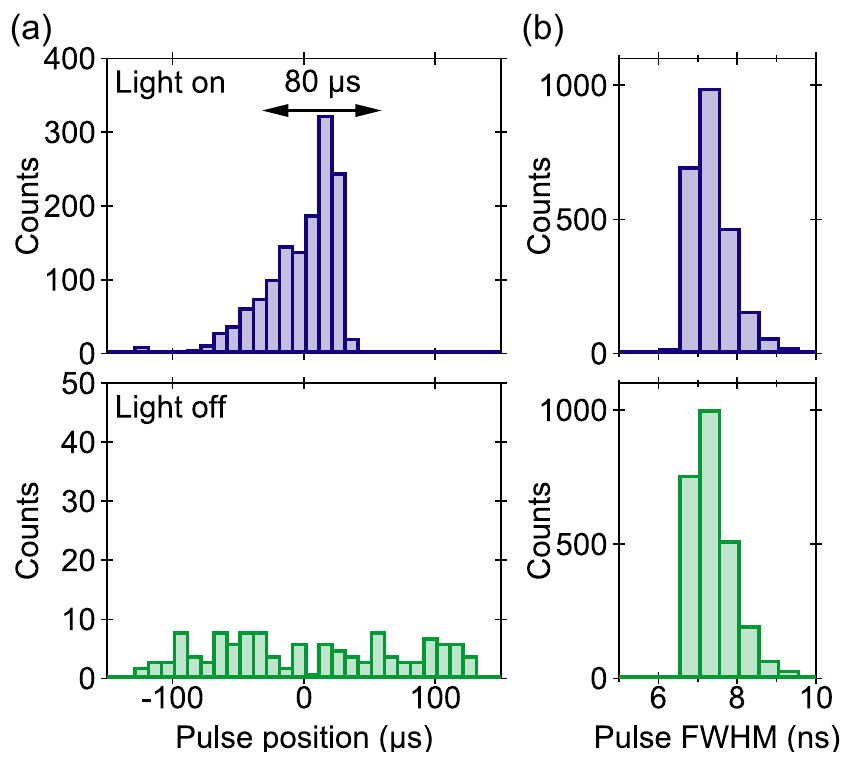}
\caption{\label{Fig: Histogram} Properties of detected photoelectrons. (a) Histogram of the photoelectron pulse positions from the center of the light pulse, when the laser is on (top panel) and off (bottom panel). The 80 $\mu\mathrm{s}$ arrow in the top panel represents the approximate width of the distribution. (b) Histogram of the FWHMs of photoelectron pulses. In the histogram, photoelectrons from other sources, detected when the light pulse was off, are also included.}
\end{figure}

We continued the detection measurement for 15 minutes; during the measurements, the photoelectron signals were sometimes observed when the light was turned off, probably due to charged particles or x-rays from other sources. 
To evaluate these background signals, we performed the same measurements with the laser turned off.
Figure \ref{Fig: Histogram}(a) shows the positions of photoelectron pulses from the center of the light pulse.
When the laser is on, the histogram is a mountain-like shape with a width of about 80 $\mu\mathrm{s}$. The result reflects the width of the light pulse and is consistent with the interpretation of wave functions as probability distributions.
On the other hand, when the laser is off, the distribution is uniform, and count rates are much less than in the case where the laser is on, which are characteristics of background signals.
The FWHMs of photoelectron signals show similar distributions in both cases with and without the light [Figure \ref{Fig: Histogram}(b)].
The wave function collapse makes the detection signals the same regardless of their origins.

We note that the distribution shape in the top panel of Figure \ref{Fig: Histogram}(a) top panel was broader than the light pulse shown in Figure \ref{Fig: Measurements}(b) left panel.
The major reason will be that the beam size was about 1 mm, larger than the size of the photo interrupter, and not negligible relative to the gap size of the chopper (about 5 mm).
In addition, the distribution was antisymmetric; the latter half was less frequent than the former half. 
We think that is because the oscilloscope took about 0.2 s to record one event; if some pulses appear sequentially within one light pulse, the oscilloscope could measure the only beginning one, and the others were discarded.
This can be also why the counts in Figure \ref{Fig: Histogram}(b) become almost the same; the oscilloscope detected the next signal immediately after recording the previous one, even without the light.

In summary, we discuss the wave function of a photoelectron emitted from the crystal surface and show that it is a cropped plane wave.
The shape of the wave function is consistent with physical intuition and the models used to describe the photoemission process and photoemission spectroscopy measurements.
We experimentally generated 80 $\mu\mathrm{s}$-long wave packets from the 4.66 eV CW laser and a single crystal of FeSe and observed them by a channel electron multiplier.
The observed signals were much narrower than the light pulse and had no difference compared to background signals due to the principle of the wave function collapse.
The photoelectron signals were distributed within a similar width of the light pulse, reflecting the wave function shape as the probability distribution.
We demonstrate that the wave function collapse, one of the most mysterious phenomena between quantum and classical physics, appears in the photoelectric effect, one of the most familiar phenomena characterizing quantum physics. 

\begin{acknowledgments}
We thank Shigeru Kasahara, Yuji Matsuda, and Takasada Shibauchi for providing FeSe samples, Yohei Kobayashi for technical advice, and Hirokazu Fujiwara, C\'edric Bareille, Takeshi Suzuki, Kozo Okazaki, and Toshiyuki Taniuchi for supporting the photoelectron detection measurements.
This work is also supported by Grants-in-Aid for JSPS Fellows (Grant No. JP21J20657) and Grants-in-Aid for Scientific Research (Grant No. JP19H00651).
\end{acknowledgments}


\end{document}